\documentclass[final,3p,times,twocolumn]{elsarticle}

\usepackage{amssymb}
\usepackage[dvipsnames]{xcolor}
\usepackage{amsmath}
\usepackage{soul}

\journal{Physics Letters B}

\newcommand{\rB}{\langle r^2 \rangle_B^{\pi^+}}
\newcommand{\rQ}{\langle r^2 \rangle_Q^{\pi^+}}

\begin{document}

\begin{frontmatter}

\title{Baryonic content of the pion}

\author[1]{Pablo Sanchez-Puertas}
\ead{psanchez@ifae.es}

\author[2]{Enrique Ruiz Arriola}
\ead{earriola@ugr.es}

\author[3,4]{Wojciech Broniowski}
\ead{Wojciech.Broniowski@ifj.edu.pl}

\affiliation[1]{organization={Institut de F\'{i}sica d'Altes Energies (IFAE) \& The Barcelona Institute of Science and Technology (BIST)},
addressline={Campus UAB},
postcode={E-08193},
city={Bellatera (Barcelona)},
country={Spain}}

\affiliation[2]{organization={Departamento de F\'{\i}sica At\'{o}mica, Molecular y Nuclear and Instituto Carlos I de  F{\'\i}sica Te\'orica y Computacional},
addressline={Universidad de Granada},
postcode={E-18071},
city={Granada},
country={Spain}}

\affiliation[3]{organization={H. Niewodnicza\'nski Institute of Nuclear Physics PAN},
postcode={31-342},
city={Cracow},
country={Poland}}

\affiliation[4]{organization={Institute of Physics},
addressline={Jan Kochanowski University},
postcode={25-406},
city={Kielce},
country={Poland}}

\begin{abstract}
The baryon form factor of charged pions arises since isospin symmetry is broken. We obtain estimates for this basic property in two phenomenological ways: from simple constituent quark models with unequal up and down quark masses, and from fitting to $e^+e^- \to \pi^+ \pi^-$ data. All our methods yield a positive $\pi^+$ baryon mean square radius of $(0.03-0.04~{\rm fm})^2$. Hence, a picture emerges where the outer region has a net baryon, and the inner region a net antibaryon density, both compensating each other such that the total baryon number is zero. For $\pi^-$ the effect is opposite.
\end{abstract}


\end{frontmatter}


 \bibliographystyle{elsarticle-num-names} 

\section{Introduction}

\begin{figure}[th]
\centering
\includegraphics[angle=0,width=0.4 \textwidth]{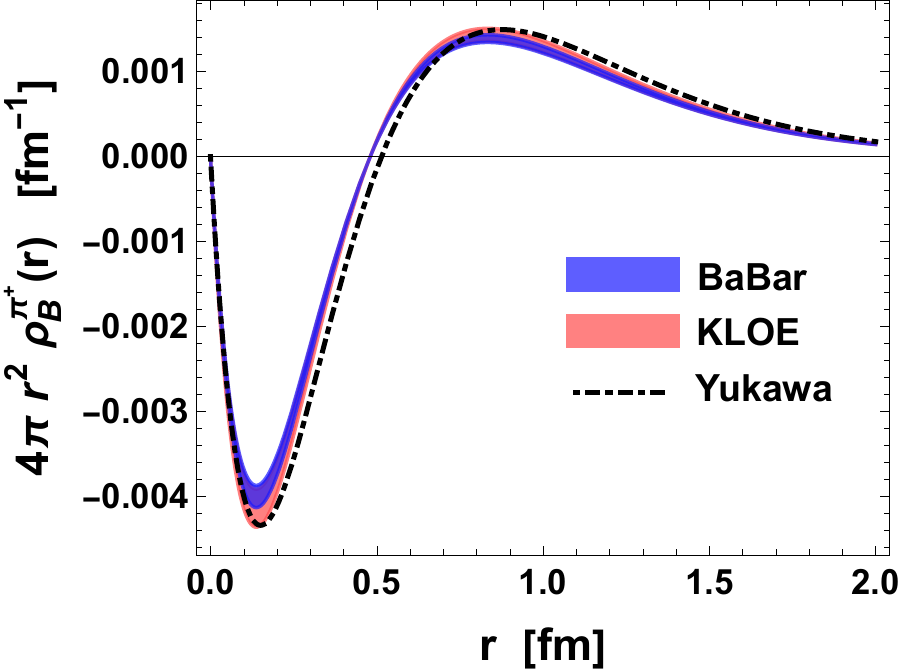}
\vspace{-2mm}
\caption{The radial baryonic charge distribution in the $\pi^+$.
The bands correspond to our extraction from 
the data, and the line to the toy Yukawa model. See the text for details. 
\label{fig:rho-rb}}
\end{figure}

Since its prediction and subsequent discovery, the pion has been
scrupulously investigated as the basic lightest hadron and the
pseudo-Goldstone boson of the dynamically broken chiral symmetry. Many
of its electroweak and mechanical properties have been studied and
determined both experimentally~\cite{Zyla:2020zbs} as well as
theoretically from a first principles point of view, with notable
recent advances from lattice QCD. In this Letter, we draw attention to
the {\em baryonic} structure of the charged pions, $\pi^+$ and
$\pi^-$, a remarkable property which, to the best of our knowledge,
has not been studied in an explicit manner before. Despite being a
zero baryon number state, the composition of the charged pion is not
baryonless.  We show that simple quark models and data analyses imply
a characteristic pattern where the matter and antimatter radial
distributions are separated at a distance of $r \sim 0.5~{\rm
  fm}$. For $\pi^+$, the inner (outer) region carries a net antibaryon
(baryon) density, and opposite for $\pi^-$
(cf.~Fig.~\ref{fig:rho-rb}).  The situation is reminiscent of the
well-known case of the electric form factor of the neutron which
carries no charge, nevertheless possesses a non-zero electric form
factor, such that (in the Breit frame) the inner (outer) region has
positive (negative) charge density, with the mean squared radius (msr)
$\langle r^2 \rangle^n_Q=-0.1161(22)~{\rm fm}^2$.  Similarly, the
neutral Kaon $K^0$ has $ \langle r^2 \rangle^{K^0 }_Q =
-0.077(10)~{\rm fm}^2$~\cite{Zyla:2020zbs} despite its null charge.
Even more striking, the nucleon is known to have a nonvanishing
strange form factor despite being
strangeless~\cite{Cohen:1993wn,Forkel:1994yx}.

\section{The baryonic form factor of the pion}

\subsection{Current conservation}

To see how the effect arises, let us first recall for
completeness some very basic facts. In QCD, one has the conservation laws
\begin{eqnarray}
\partial_{\mu} \left[ \bar q_a (x) \gamma^\mu q_b (x) \right]= i (m_a-m_b) \bar q_a(x)  q_b (x),  
\end{eqnarray}
with $q_{j}(x)$ denoting the quark field with $N_c$ colors and flavor
$j=u,d,s,c,b,t$, which for equal quark masses corresponds to the conservation of the vector current 
and ensures the quark number conservation for any species. For
the specific case of the pion, we neglect $s$ and
heavier flavors, as they represent corrections suppressed by the Okubo-Zweig-Iizuka (OZI)  
rule and are subleading in the large-$N_c$ limit of QCD. 
In this case the baryon current (isosinglet) and the third isospin component of the isovector current are
\begin{equation}
J_B^\mu = \frac1{N_c} \left(\bar u \gamma^\mu u + \bar d \gamma^\mu d \right) ,
 \ \
J_3^\mu = \frac1{2} \left (\bar u \gamma^\mu u - \bar d \gamma^\mu d \right), 
\end{equation}
where $N_c\equiv 3$ is assumed in the following. 
With these definitions, the Gell-Mann--Nishijima formula provides the electromagnetic
current $J_{Q}^{\mu} = J_{3}^{\mu} + \frac{1}{2}J_{B}^{\mu} $.  The baryon, isospin,
and charge form factors are defined via the on-shell matrix elements of the corresponding currents, namely
\begin{equation}
\langle \pi^a(p) \mid J_{B,3,Q}^\mu(0) \mid \pi^a(p+q)\rangle = (2p^\mu +q^\mu) F^a_{B,3,Q}(q^2),
\end{equation}
with $F_{Q}^a(q^2)=F_3^a(q^2)+\frac{1}{2}F_B^a(q^2)$. 

\subsection{Charge conjugation and Isospin violation}

Now come the standard symmetry arguments. 
Since $J_B^\mu$ is odd under charge conjugation $C$, it implies that for the $C$-even neutral pion $F^{\pi^0}_B(q^2)=0$ identically, while for the charged pions it provides the relation $F^{\pi^+}_B(q^2) = -F^{\pi^-}_B(q^2)$.
Similarly, for the case of an exact isospin symmetry,
i.e., with $m_u=m_d$ and neglecting small electromagnetic effects, G-parity symmetry yields $F^{\pi^\pm}_B(q^2)=0$.  However, this is no longer the
case in the real world where the isospin is broken with
$m_d>m_u$. G-parity ceases to be a good symmetry and 
$F^{\pi^\pm}_B(q^2)$ may be --- and in fact is --- non-zero, with 
$F^{\pi^+}_B(q^2)=-F^{\pi^-}_B(q^2)\neq 0$. Moreover, if we take (say, for $\pi^+$) $Z \bar d i \gamma_5 u$ as an interpolating field,
then a direct application of the Ward-Takahashi-Green identities for the
conserved $B$ and $Q$ currents implies (for the canonical pion field) $F^{\pi^\pm}_B(0)=0$, $F^{\pi^\pm}_Q (0)=\pm 1$.

\subsection{Coordinate space interpretation}

A popular interpretation of form factors is based on choosing the
Breit reference frame, where there is no energy
transfer. Then the form factor in the space-like region $q^2 = - \vec q^2 \equiv -Q^2 \le 0$ allows one to construct the
(naive) 3-dimensional baryon density as
\begin{eqnarray}
\rho_B(r) = \int \frac{d^3 q}{(2\pi)^3} e^{i \vec q \cdot \vec r}
F_B(-\vec q^2 ). \label{eq:ft}
\end{eqnarray}
As ambiguities stemming from relativity arise~\cite{Jaffe:2020ebz,Soper:1976jc}, it has been
argued that a frame-independent interpretation can be formulated in terms of a  transverse
density in the 2-dimensional impact-parameter $\vec{b}$~\cite{Burkardt:2000za}, where 
instead of Eq.~(\ref{eq:ft}) one takes the Fourier integral with $\exp({i \, \vec q_\perp \cdot \vec b})$.
(see, e.g., \cite{Miller:2010nz} for a review and Ref.~\cite{Lorce:2020onh}). Here, for our
illustrative purpose, we choose to show the $r$-space densities, as the
$b$-space results are simply related and qualitatively the same.

We have no obvious sources coupling the charged pions solely to the
baryon current, hence a direct experimental measurement of
$F_B^{\pi^\pm}(q^2)$ is not possible. This is also in common with the
neutron electric form factor, where a direct determination is hampered
by the absence of free neutron targets and its extraction requires
scattering on bound neutrons in the deuteron.\footnote{
  Remarkably, the earliest determinations of the neutron radius were
  promoted by Fermi and Marshall in 1947~\cite{Fermi:1947zz} from
  looking at ultracold neutron - atom scattering transmission
  experiments. These involve the total but not the differential
  cross section (see also PDG~\cite{Zyla:2020zbs}), hence precluding a
  form factor determination.} The analysis requires an accurate
deuteron wave function as well as meson exchange current
effects~\cite{Gilman:2001yh}, hence the need for additional
theoretical input. Returning to the novel case of the so far
disregarded pion baryonic form factor, we will content ourselves with
rather unsophisticated but complementary and realistic estimates.
They are based on dimensional analysis, quark models, and an
extraction from $e^+ e^- \to \pi^+ \pi^-$ 
data analysis. Within uncertainties, our results are consistent.

\section{Model estimates}

\subsection{Dimensional analysis}

A generic order of magnitude estimate of the discussed isospin
violating effect can be obtained at the leading order in the pion
momenta and the quark mass splitting $\Delta m \equiv
{m_d-m_u}=2.8(2)$~MeV (in this work we use $m_u = 2.01(14)$~MeV and
$m_d = 4.79(16)$~MeV~\cite{Davies:2009ih}).  We expect the two-pion
contribution to the baryon current to be of the effective form 
\begin{equation}
J_B^\mu = -2i \frac{c \Delta m}{ \Lambda^3} \partial_\nu \left( \partial^\mu \pi^+ \partial^\nu \pi^-
-\partial^\nu \pi^+ \partial^\mu \pi^- \right) + \dots,  \label{eq:chir}
\end{equation}
with $c$ an undetermined dimensionless number and $\Lambda$ a typical
low energy hadronic scale (say, $m_\rho \sim 770$~MeV). As it should,
this current is odd under $C$, is trivially conserved, and its
contribution vanishes for $q^2=0$, providing the form factor $
F_B^{\pi^+}(q^2) = q^2 c \Delta m /\Lambda^3 + \dots$, with msr $\rB{}
= 6 c \Delta m /m_\rho^3 \simeq c~0.002~{\rm fm}^2 \simeq c (0.04~{\rm
  fm})^2$, a small number compared to the electric charge radius
$\rQ{} = (0.659(4)~{\rm fm})^2 = 0.434(5)~{\rm
  fm}^2$~\cite{Zyla:2020zbs}.  
  
One may seek further guidance in Chiral Perturbation Theory.
In particular, the term in Eq.~(\ref{eq:chir}) arises starting from the ${\cal
    O}(p^6)$ chiral Lagrangian~\cite{Bijnens:1999sh},
\begin{equation}
\mathcal{L}_{63+65}^{(6)} =\frac{i}{F^2}\langle f_{+\mu\nu}\left( C_{63}\{ \chi_+,u^{\mu}u^{\nu}\} 
+ C_{65}u^{\mu}\chi_+u^{\nu}\right) \rangle, 
\end{equation}
from where we find an explicit relation $c/\Lambda^3 = \frac{8B_0}{N_c  F^4}(2C_{63} -C_{65})$. It involves
two $C_i$ coefficients, for which currently there 
are no independent estimates.
The naturalness condition yields $C_i F_{\pi}^{-4} \sim m_{\rho}^{-4}$, hence $c\sim 1$, as previously argued.

\subsection{Yukawa quark model}

\begin{figure}[b]
\centering \includegraphics[angle=0,width=0.37
  \textwidth]{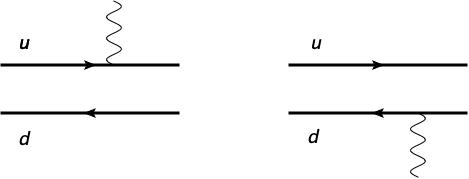}
\vspace{-2mm}
\caption{Diagrams for evaluating the form factors of  $\pi^+$ in the impulse approximation. With postulated Yukawa wave functions we get the VMD model in the isovector channel.}
\label{fig:impulse}
\end{figure}

\begin{figure}
\centering
\includegraphics[angle=0,width=0.4 \textwidth]{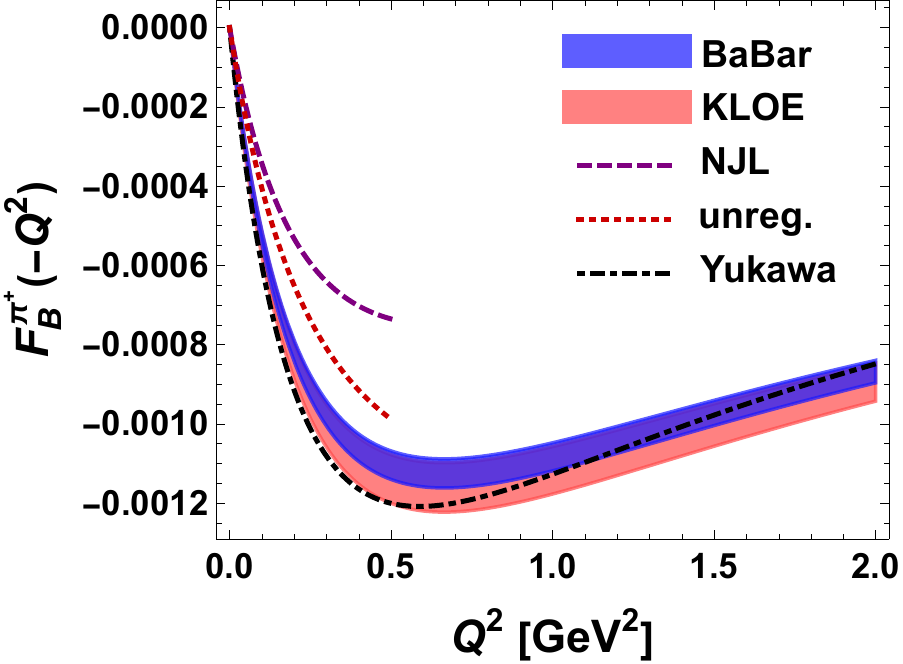}
\vspace{-2mm}
\caption{The baryon form factor of $\pi^+$ in the space-like region. The bands correspond to the extraction from 
the data, and the lines to various models described in the text. \label{fig:Ft} }
\end{figure}

Next, we come to our quark model estimates for ${\Delta m \neq 0}$
effects. To start, we explore the fact that the coordinate
representation motivates a toy constituent-quark model based on the
familiar {\em impulse approximation} in nuclear physics (cf.~Fig.~\ref{fig:impulse}).
In this framework $\rho_B(r)= B_{u}|\Psi_{u}(\vec x)|^2 + B_{\bar d}
|\Psi_{\bar d}(\vec x)|^2$, where the quark baryon numbers are
$B_{u}=-B_{\bar d}= \frac{1}{N_c}$.  Motivated by Vector Meson
Dominance model (VMD), that provides a reasonable description for the
isovector channel, $F_3^{\pi^+}(q^2)=M_{\rho}^2/(M_{\rho}^2 -q^2)$
(and allows for a simple extension to other pseudoscalar mesons), we
take the normalized Yukawa-like probabilities
%
  $|\Psi_{i}(\vec x)|^2= M_i^2 e^{-2 M_i r}/(\pi r)$, with $M_{u,d}=M
\mp \frac{1}{2}\Delta m$ and $M$ denoting the constituent quark mass.
This ensures that, for $\Delta m=0$, the resulting isovector and
charge form factors $F_{Q,3}^{\pi^+}(-Q^2)=4M^2/(4M^2+Q^2)$ reproduce
the VMD phenomenology provided we take $M=\frac{1}{2}m_\rho \simeq
385~{\rm MeV}$, while for the baryon form factor we find 
\begin{eqnarray}
F_B^{\pi^+} (-Q^2) = \frac{1}{N_c} \left[ \frac{4M_u^2}{4M_u^2+Q^2} -\frac{4M_d^2}{4M_d^2+Q^2} \right]. \label{eq:yuk}
\end{eqnarray} 
Eq.~(\ref{eq:yuk}) yields
$\langle r^2 \rangle_B^{\pi^+ } \simeq 3 \Delta m/N_c M^3 \simeq (0.04~{\rm fm})^2$.
Our result for the baryon density is plotted in Fig.~\ref{fig:rho-rb},
where we note a  change of sign 
of the baryon density at
$r_0=\log(M_d/M_u))/(M_d-M_u) \simeq 1/M \simeq 0.5$~fm. The corresponding form factor is shown in Fig.~\ref{fig:Ft}. It takes a
minimum $F_{\rm min} = -\Delta m/2M N_c \simeq -0.0012$ at $Q^2 \simeq 4M^2 \simeq 0.5~{\rm GeV}^2$.

Generalizing to other meson states, such as the Kaons, $D$, or $B$ mesons, with
heavy-light  constituents, yields a shift of the crossing point to shorter distances. 
With $M_a=M+m_a$, and taking
$m_s=100~{\rm MeV}$, $m_c=1.27~{\rm GeV}$, $m_b=4.65~{\rm GeV}$, the toy model 
gives $\langle r^2 \rangle_B^{K^{+,0}}= (0.22~{\rm fm})^2 $, $\langle r^2 \rangle_B^{D^{0,-}}= (0.35~{\rm fm})^2 $, and
$\langle r^2 \rangle_B^{B^{+,0}}= (0.36~{\rm fm})^2 $, with equal and negative values for the corresponding antiparticles.

\begin{figure}[b]
\centering
\includegraphics[angle=0,width=0.23 \textwidth]{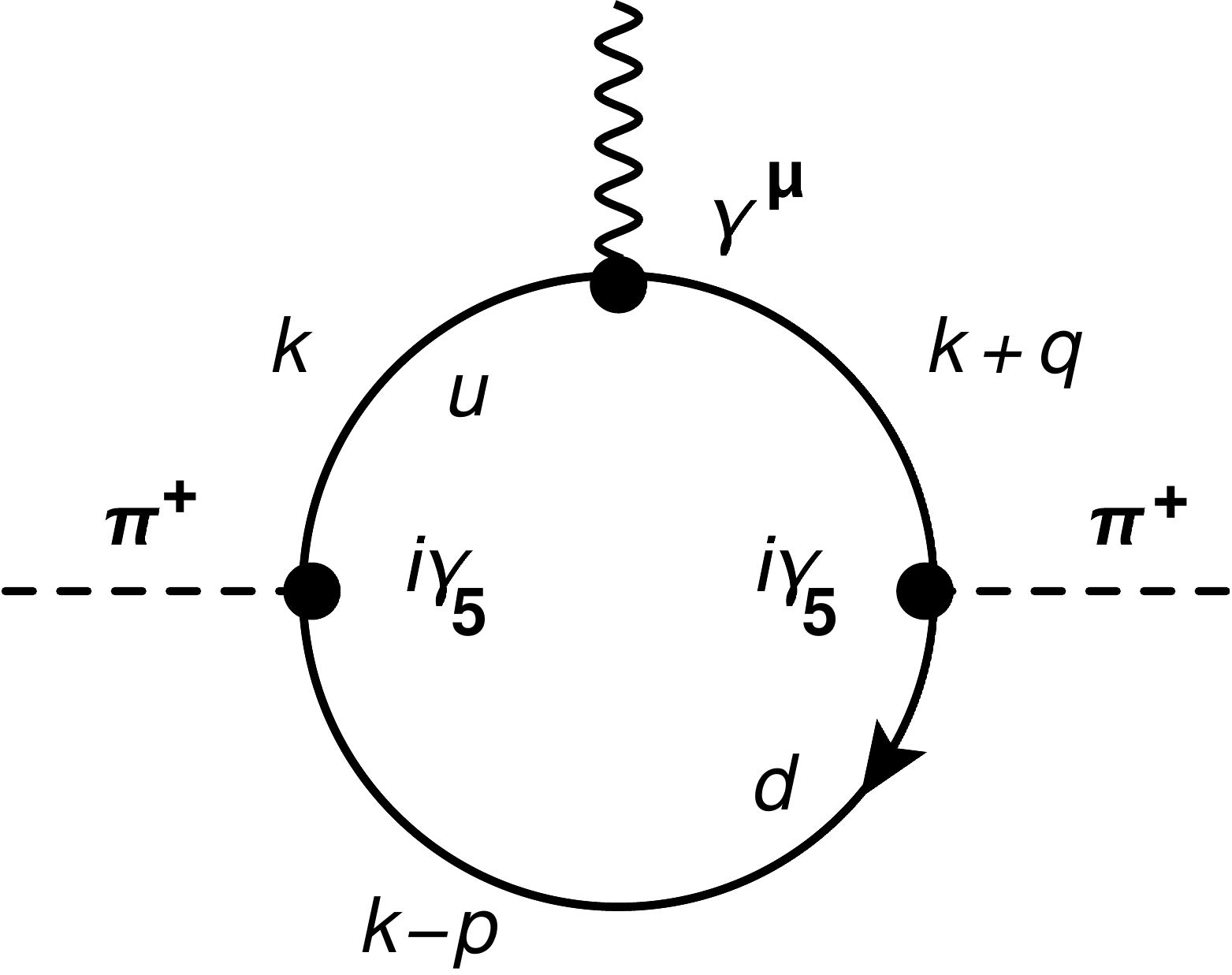} \includegraphics[angle=0,width=0.23 \textwidth]{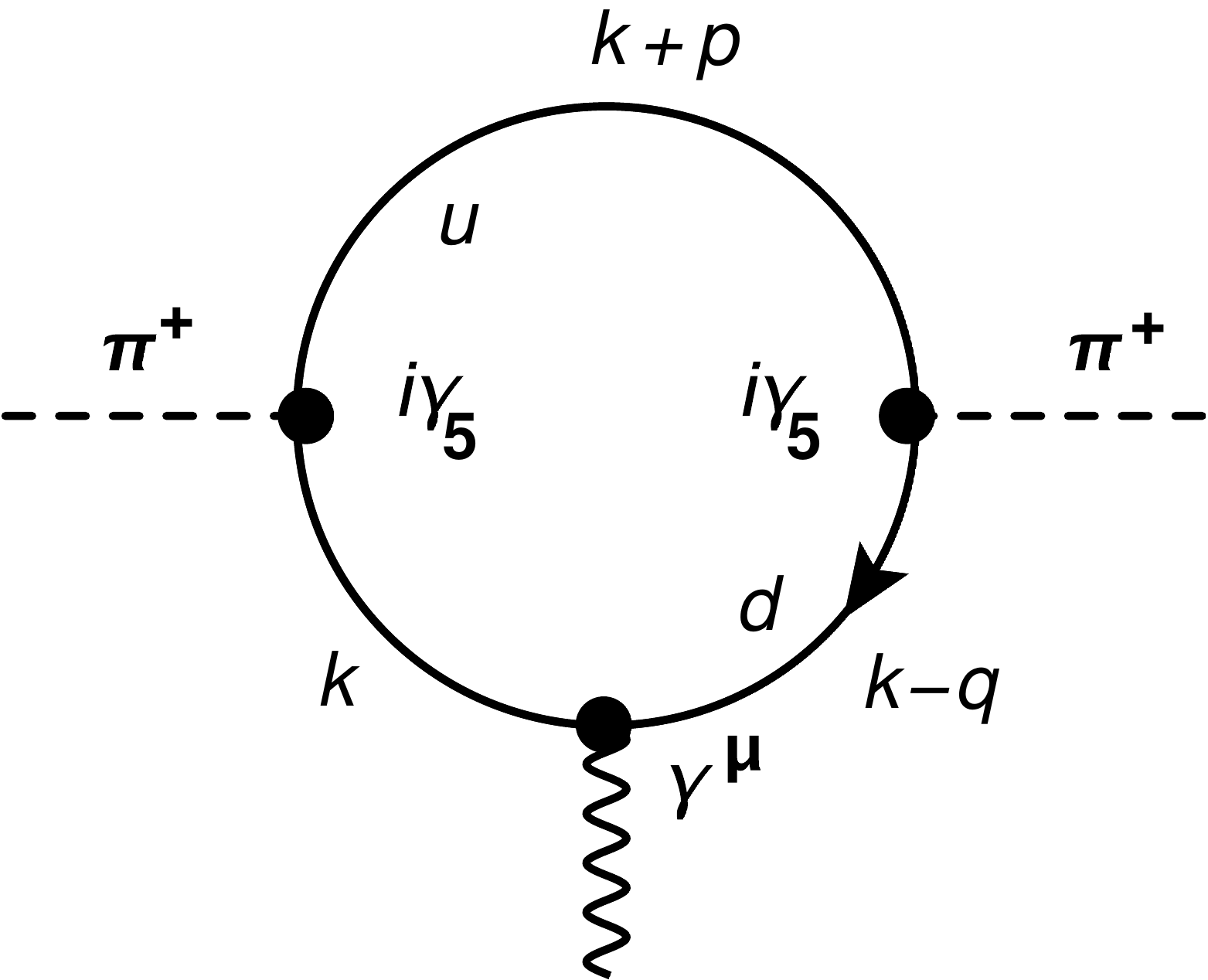}
\vspace{-2mm}
\caption{Feynman diagrams for evaluating the baryon form factor of $\pi^+$ in the NJL model. \label{fig:feyn}}
\end{figure}

\subsection{Chiral quark model}

We now pass to a quark model where the pion is described with a fully
relativistic $\bar q q$ dynamics. The Nambu-Jona--Lasinio (NJL) model
with consitutent quarks (see \cite{RuizArriola:2002wr} and Refs. therein) 
implements the spontaneously broken chiral symmetry, providing the
quarks dynamically with a constituent quark mass $M$. A leading-$N_c$
diagrammatic representation of the baryon form factor is given in
Fig.~\ref{fig:feyn}, where we indicate the momenta,
and the $\pi^+ u \bar{d}$ coupling constant of the point-like coupling is $(M_u+M_d)/\sqrt{2}F_{\pi}$, with $F_{\pi}$ 
denoting the pion weak decay constant. The Lorentz, gauge, and chiral symmetries
are preserved by implementing the Pauli-Villars regularization~\cite{RuizArriola:2002wr}, imposed in these effective models to suppress the hard momentum contribution from the quark loop.

The result of a standard evaluation has the compact form for the expressions leading in the quark mass splitting $\Delta m$:
\begin{eqnarray}
&& F^{\pi^+}_Q(t)= \frac{4N_c M^2}{F_{\pi}^2} I_1(t) \nonumber \\
&& F^{\pi^+}_B(t)=  \frac{16 M^3 \Delta m }{F_{\pi}^2} \left [J(t) -I_2(t) \right], 
\end{eqnarray}
with the basic one-loop integral evaluated in the Euclidean space as
\begin{eqnarray}
&& I_p(t)= -i \!\!\int \!\!\frac{d^4k}{(2\pi)^4} G(k)^p G(k+q), \nonumber \\
&& J(t)=-i \!\! \int \!\!\frac{d^4k}{(2\pi)^4}  G(k)G(k+q)G(k-p),
\end{eqnarray}
where the quark propagator is $G(l)\equiv 1/(l^2 - M^2+i \epsilon)$.
The subscript `reg' indicates the regularization, imposed in these effective models to suppres the hard momentum contribution from the loop. Explicit 
calculation in the chiral limit yields the result
\begin{eqnarray}
~\hspace{-9mm} && F^{\pi^+}_Q(t)=\left . 1+\frac{M^2 N_c \left(2 s-\log \left(\frac{s+1}{1-s}\right)\right)}{4 \pi ^2 f^2 s} \right |_{\rm reg},  \\
\hspace{-9mm}&& F^{\pi^+}_B(t)=\left . \frac{\Delta m  M^3 \!\left[ \frac{1}{2}\log^2 \left(\frac{s+1}{1-s}\right)
+ \log \left(\frac{s+1}{1-s}\right)\right]}{\pi ^2 F_{\pi}^2 t}\right |_{\rm reg}\!\!\!\!, \nonumber 
\end{eqnarray}
with the short-hand notation \mbox{$s=1/\sqrt{1-4M^2/t}$} introduced.
The  low-$Q^2$ expansion is 
\begin{eqnarray}
&& F^{\pi^+}_Q(-Q^2)=\left . 1-\frac{N_c Q^2}{24 \pi ^2 F_{\pi}^2}+\dots  \right |_{\rm reg},  \\
&& F^{\pi^+}_B(-Q^2)=\left . - \frac{\Delta m  Q^2}{24 \pi ^2 F_{\pi}^2 M}+\dots  \right |_{\rm reg}.  \nonumber
\end{eqnarray}
The baryon radius is $\rB{} = (0.03~{\rm fm})^2$
with $M=0.3-0.35$~GeV and the Pauli-Villars cut-off $\Lambda \simeq 0.7$~GeV, adjusted to give the physical value of $F_{\pi}$.  Actually,
since the one-loop calculation of Fig.~\ref{fig:feyn} yields a finite result
for the electric charge msr~\cite{Tarrach:1979ta} and for the baryon
msr, without regularization we obtain a numerically very similar (though not identical) value, $\langle r^2 \rangle_B^{\pi^+}=
({\Delta m}/N_c M) \langle r^2 \rangle_Q^{\pi^+} \sim (0.03~{\rm fm})^2$.  The baryon form factor in the NJL model with PV regularization (NJL) and in the
unregularized case (unreg.) are plotted in Fig.~\ref{fig:Ft} for $Q^2 \lesssim \Lambda^2 \sim 0.5~{\rm GeV}^2$. Momenta higher than the cut-off are hard and are outside of the fiducial range of the effective quark models, hence the functions are not plotted there. 

\subsection{Vector meson dominance}

Alternatively to quark models, 
on the basis of the quark-hadron duality we may adopt a purely hadronic
description that shall illustrate the significance of the  $\rho - \omega$ 
mixing in the context of the baryon form factor. In VMD, the
physical $\omega$ and $\rho^0$ mesons are linear combinations of the
isoscalar $\omega^0$ and isovector $\rho^3$ states with a mixing
angle $\theta$. With the  current-field
identities~\cite{Kroll:1967it,deAlfaro:1973zz} and the matrix
elements $\langle 0 |J_B^\mu |\omega^0 \rangle = \frac{1}{N_c}f_B \epsilon^\mu $ and 
$\langle 0 |J_3^\mu |\rho^3 \rangle = \frac{1}{2} f_3 \epsilon_\mu $, 
the form factors in the space-like region read
\begin{eqnarray}
F_3(-Q^2)  &=& \frac{f_3}2 \left[ \frac{\sin \theta g_{\omega \pi \pi}}{Q^2+m_\omega^2} + \frac{\cos \theta g_{\rho\pi\pi}}{Q^2+m_\rho^2} \right] \nonumber \\
F_B (-Q^2) &=& \frac{f_B}{N_c} \left[ \frac{\cos \theta g_{\omega \pi \pi}}{Q^2+m_\omega^2} - \frac{\sin \theta g_{\rho\pi\pi}}{Q^2+m_\rho^2} \right] 
\end{eqnarray}
with $g_{\omega\pi\pi}$ and $g_{\rho\pi\pi}$  the couplings for 
$\omega \to \pi^+ \pi^-$ and $\rho \to \pi^+ \pi^-$ decays and $f_{B,3}$ related to $\rho/\omega\to\ell^+\ell^-$ decays (see for instance Ref.~\cite{Hanhart:2016pcd}).  
The conditions $F_B(0)=0$ and $F_3(0)=1$ imply
$g_{\rho\pi\pi}=2 m_\rho^2 \cos \theta /f_3$, $g_{\omega\pi\pi}=2 m_\omega^2 \sin \theta /f_3$, and
\begin{eqnarray}
&& F_3 (-Q^2) = \frac{\cos^2 \theta m_\rho^2}{Q^2+m_\rho^2}+ \frac{\sin^2 \theta m_\omega^2}{Q^2+m_\omega^2}, \nonumber \\ 
&& F_B (-Q^2) =\frac{ Q^2 f_B\sin (2 \theta) (m_\omega^2-m_\rho^2)}{ f_3 N_c (Q^2+m_\rho^2)(Q^2+m_\omega^2)}. \label{eq:fQ-vmd} 
\end{eqnarray} 
The above formula nicely illustrates basic physical features: the association of emergence of $F_B (-Q^2)$ with the $\rho-\omega$ mixing, and its vanishing value at $Q^2=0$.
    

    However, Eqs.~(\ref{eq:fQ-vmd}) hold literally for narrow-width 
    mesons only, which is certainly not the case for the broad $\rho$ 
    resonance and precludes building a successful phenomenology. 
    For that reason we do not elaborate numerically Eqs.~(\ref{eq:fQ-vmd}), 
    treating them only as a guideline for a more sophisticated analysis 
    of the next section, where the width of resonances is properly 
    incorporated.


\section{Data analysis}

We now use the available high statistics data in the time-like region to extract the baryon form factor. Actually, the
Gell-Mann--Nishijima formula for the form factors
would allow for a direct determination if it were not for the
fact that, unlike for $|F_{Q}^{\pi^\pm}(q^2)|$ accessible from the
$e^+e^- \to \pi^+ \pi^-$ reaction from BaBar~\cite{Aubert:2009ad} and
KLOE~\cite{Aloisio:2004bu,Ambrosino:2008aa,Ambrosino:2010bv,Anastasi:2017eio}), the
$F_3^{\pi^\pm}(q^2)$ form factor remains unknown.  One could in principle
consider the flavor-changing current $J_\mu^+ = \bar u \gamma_\mu d$
and the corresponding form factor $F_+(q^2)$ appearing in the matrix
element $ \langle \pi^0 | J_\mu^+ | \pi^+ \rangle $, 
determined to a high precision in $\tau \to \pi^+ \pi^0 \nu_\tau$ decays
by Belle~\cite{Fujikawa:2008ma}.  In the strict isospin limit ($\Delta
m=0$), the latter is simply related to $F_3^{\pi^\pm}(q^2)$ via isospin
rotation, a relation that has been exploited in the context of the muon
$(g-2)$~\cite{Dung:1996rp,Cirigliano:2001er,Cirigliano:2002pv,Ghozzi:2003yn,%
Benayoun:2007cu,Maltman:2005qq,Wolfe:2009ts,Davier:2009ag,Wolfe:2010gf,Jegerlehner:2011ti,Miranda:2020wdg},
but without paying an effort to extract $F_B^{\pi^\pm}(q^2)$. Moreover, the isospin relation
no longer holds if $\Delta m\neq 0$ or the electromagnetic effects are
accounted for.  Actually, while the isospin version of the
Ademollo-Gatto non-renormalization theorem~\cite{Ademollo:1964sr}
implies $F_3^{\pi^{\pm}}(0)=F_+ (0) + \mathcal{O}(\Delta m^2)$, this
is no longer true at finite momentum transfer, where
$F_3^{\pi^{\pm}}(q^2)=F_+ (q^2) + \mathcal{O}(\Delta m)$, comparable
itself to the effect we aim to extract,
$F_B^{\pi^{\pm}}(q^2)= \mathcal{O}(\Delta m)$.  Indeed, our attempts
to do so with the aid of dispersion relations 
provided noisy results.  These can be ascribed 
to the isospin violating corrections, as we detail in the analysis below.

\begin{figure}[h!]
\centering
\includegraphics[angle=0,width=0.4 \textwidth]{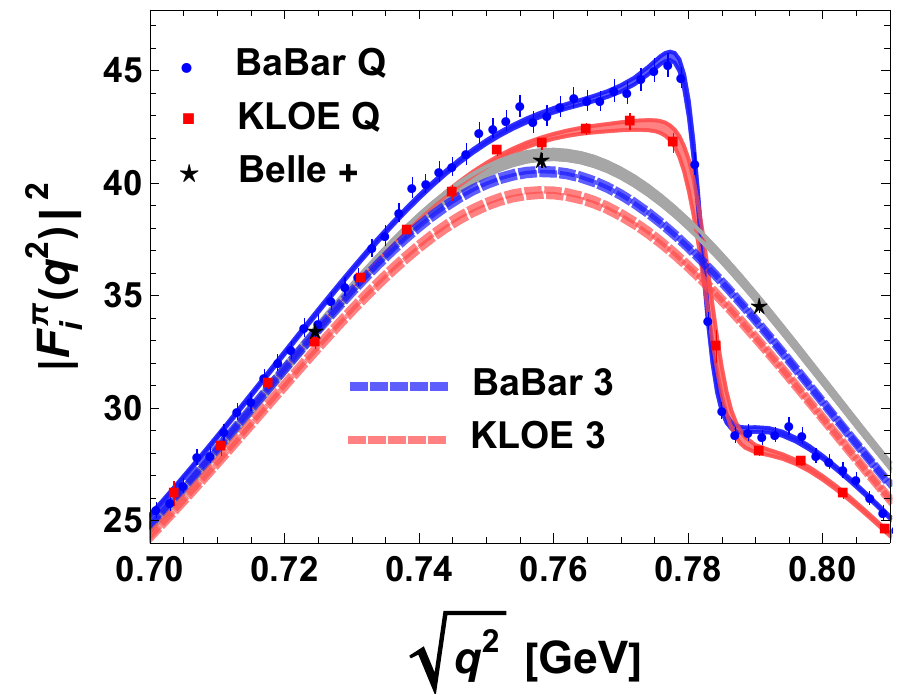}
\vspace{-2mm}
\caption{Various form factors in the $\rho-\omega$ region. The points indicate the experimental data
from  $e^+e^- \to \pi^+ \pi^- $ (BaBar and KLOE for the $Q$ form factor), and 
from $\tau^- \to \pi^0 \pi^- \nu_\tau$ (Belle for the $+$ form factor).
Our fits are represented with the solid bands following the points, where the widths reflect the statistical errors.
We also give the resulting fits to the $3$ form factors, indicated with the dashed bands. \label{fig:compare-rho-omeg}}
\end{figure}

For the above reasons we restrict ourselves to the model 
decomposition~\cite{Cirigliano:2002pv}
\begin{equation}
F_{Q}^{\pi^{\pm}}(q^2) = F_3 (m_{\pi^+} , m_{\pi^-}, m_{\rho^0} ,q^2) + \frac{1}{2}F_{B}^{\pi^{\pm}} (q^2), \label{eq:ff-fit}
\end{equation}
to extract $F_{B}^{\pi^{\pm}}(q^2)$. As a consistency check of the 
approach, we also fit $F_+(q^2) = F_3 (m_{\pi^+} , m_{\pi^0}, m_{\rho^+} ,q^2)$. We adopt 
the Gounaris-Sakurai~\cite{Gounaris:1968mw} parametrization used in the BaBar 
analysis~\cite{Lees:2012cj} for $F_3(q^2)$, whereas the $\rho-\omega$ mixing term is taken in the form
\begin{equation}\label{eq:rhoomega}
\frac{1}{2}F_B^{\pi^{\pm}}(q^2) =  c_{\rho\omega} q^2 D_{\rho}(q^2)D_{\omega}(q^2),
\end{equation} 
ensuring that $F_B^{\pi^{\pm}}(0)=0$. Further, $c_{\rho\omega}$ is real, not to spoil analyticity (see
Eq.~(\ref{eq:dr}) below) by a constant non-zero phase. We emphasize these features were not implemented in the BaBar analysis, while the form of
$D^{-1}_V(t)= \tilde{m}_V^2[ m_V^2 - t - i m_V \Gamma_V(t)]^{-1}$ is taken as in the BaBar
parametrization~\cite{Lees:2012cj}.

The results are shown in Fig.~\ref{fig:compare-rho-omeg}.
The two highest-reaching bands correspond to our fits
to $F_{Q}^{\pi^\pm}(q^2)$ from BaBar~\cite{Aubert:2009ad} and
KLOE~\cite{Aloisio:2004bu,Ambrosino:2008aa,Ambrosino:2010bv,Anastasi:2017eio} 
with QED effects from vacuum polarization and final state radiation removed. 
The corresponding $F_3^{\pi^\pm}(q^2)$ form factors resulting from these fits are 
shown as the two dashed bands. These can be compared 
to the fit to the $F_+(q^2)$ form factor from Belle, shown 
as a band. The mild difference 
is clear and can be ascribed to the mentioned
isospin-breaking corrections. 
The obtained value of the mixing parameter from the BaBar/KLOE data is $c_{\rho\omega} =[36(1)/37(2)]\times10^{-4}~\textrm{GeV}^{-2}$.

To extract the behavior in the spacelike region, we make use of
analyticity and the perturbative high-energy behavior at large $q^2$,
$F_B^{\pi^{\pm}}(q^2) = {\cal O}(1/q^2)$~\cite{Lepage:1980fj}, that
allows one to write down the unsubtracted (and subtracted) dispersion
relations
\begin{eqnarray}
\hspace{-8mm} F_B^{\pi^{\pm}} (q^2)= \frac1{\pi} \int_{4 m_{\pi^+}^2}^\infty \!\!\!\!\!\!\! ds \frac{{\rm Im} F_B^{\pi^{\pm}} (s)}{s-q^2} = 
\frac{q^2}{\pi} \int_{4 m_{\pi^+}^2}^\infty \!\!\!\!\!\!\! ds \frac{{\rm Im} F_B^{\pi^{\pm}} (s)}{s(s-q^2)}.   \label{eq:dr}
\end{eqnarray}

The results of Fig.~\ref{fig:Ft} show that despite a discrepancy between BaBar and
KLOE in the time-like region (cf. Fig~\ref{fig:compare-rho-omeg}), the  
obtained baryonic form factors in the
space-like region are compatible. 
Likewise, the baryonic radius computed from Eq.~(\ref{eq:dr}) 
%
%
yields $\langle r^2 \rangle_B^{\pi^+}= \left( 0.0411(7)~{\rm
fm} \right)^2$ for BaBar and $\left( 0.0412(12)~{\rm fm} \right)^2$
for KLOE, are in a remarkable agreement.  

\section{Conclusions and outlook}

We summarize in
Table~\ref{tab:sum} all the obtained estimates for the baryonic msr of
the charged pion, which fall in the range $\langle r^2 \rangle_B^{\pi^+}= ((0.03-0.04)~{\rm fm})^2 =(0.001-0.002)~\textrm{fm}^2 $. The agreement with the Yukawa model 
complies to the natural understanding on the positive sign of
the radius for $\pi^+ (u \bar d)$, where the lighter $u$ component
is more extended than the heavier $\bar d$ component.
\begin{table}[t]
\caption{Various estimates for the baryonic radius of $\pi^+$. \label{tab:sum}}
\footnotesize
\begin{tabular}{lll}\hline
approach &  $\langle r^2 \rangle_B^{\pi^+}$  & comment \\ \hline 
effective Lagrangian & $ c (0.04~{\rm fm})^2$ & $c$ - number of order 1 \\
toy Yukawa model   & $ (0.04~{\rm fm})^2$ & \\
NJL with PV reg. &  $ (0.03~{\rm fm})^2$ & \\
NJL without reg. &  $ (0.03~{\rm fm})^2$ & \\
\hline
BaBar & $ (0.041(1)~{\rm fm})^2$ & exp. stat. error only \\
KLOE & $ (0.041(1)~{\rm fm})^2$ & \\ \hline
\end{tabular}
\end{table}
Comparing the estimates of Table~\ref{tab:sum} to the accuracy of the experimental charge radius,
$\langle r^2 \rangle_Q^{\pi} = (0.659(4)~{\rm fm})^2 =0.434(5)~{\rm fm}^2$~\cite{Zyla:2020zbs}, or to the most recent {\it ab initio} lattice
QCD calculations with physical averaged quark masses, 
$\langle r^2 \rangle_Q^{\pi} = (0.648(15)~{\rm fm})^2 =0.42(2)~{\rm fm}^2$~\cite{Gao:2021xsm}
and $\langle r^2 \rangle_Q^{\pi} =0.430(5)(12)~{\rm fm}^2$~\cite{Wang:2020nbf},
 we note that the  
signal is 
a factor of $4 -10$ too small to affect current charge radius determinations.
However, the small baryonic msr is also coming from the $1/N_c$ prefactor preceding the baryon current. Without this factor, one finds $N_c\langle r^2 \rangle_B^{\pi^+} = \langle r^2 \rangle_u^{\pi^+} - \langle r^2 \rangle_{\bar{d}}^{\pi^+}= (0.003-0.005)~\textrm{fm}^2$, that might be within reach in lattice QCD.

The case of the baryonic content of the Kaon
is much more promising. Since for $\bar{d}$ and $s$ quarks the baryon
number equals minus the charge, in models with structureless
constituent quarks (such as our toy Yukawa model) $ \langle r^2
\rangle^{K^0 }_B = -\langle r^2 \rangle^{K^0}_Q $. The Yukawa model
yields $\langle r^2 \rangle^{K^0 }_B \simeq (0.22 ~{\rm fm })^2 \simeq 0.05~{\rm fm}^2 $, whereas
PDG~\cite{Zyla:2020zbs} quotes $ \langle r^2 \rangle^{K^0 }_Q =
-(0.28(2)~{\rm fm})^2 = -0.077(10) ~{\rm fm}^2 $, with the uncertainty
5 times smaller than the Yukawa model estimate for the baryon msr.
Similarly, VMD predicts $F_B^{K^0}(q^2) = N_c^{-1}(D_{\omega}(q^2)
-D_{\phi}(q^2))$, thus $\langle r^2 \rangle^{K^0 }_B =
(0.23~\textrm{fm})^2 = 0.052~\textrm{fm}^2$, while a data-driven
analysis such as that in Sect.~4 would be more involved. In
particular, the $K\bar{K}$ threshold, well above the $\rho,\omega$
mesons, would require a more elaborated analysis as outlined in
Refs.~\cite{BaBar:2013jqz,BaBar:2014uwz}. 

Hopefully, more accurate experiments and
their corresponding analyses of the vector form factors would 
provide a better understanding of the fundamental issue of the matter-antimatter distribution in 
pseudoscalar mesons.

Supported by the European H2020 MSCA-COFUND (grant No. 754510) 
and H2020-INFRAIA-2018-1 (grant No. 824093), 
the Spanish MINECO (grants FPA2017-86989-P and SEV-2016-0588), 
and Generalitat de Catalunya (grant 2017SGR1069) (PSP), 
the Spanish MINECO and European FEDER funds
(grant FIS2017-85053-C2-1-P), 
Junta de Andaluc{\'i}a (grant FQM-225) (ERA), 
and the Polish National Science Centre grant
2018/31/B/ST2/01022 (WB).





\bibliography{refs-barpi}

\end{document}